\documentclass[12pt]{article}
\usepackage{amssymb}
\usepackage{euscript}

\usepackage[latin1]{inputenc}
\usepackage[T1]{fontenc}
\usepackage{amsmath}
\usepackage{amsfonts}
\usepackage{amssymb}
\usepackage{latexsym}
\usepackage[english]{babel}
\usepackage[dvips]{graphicx,color}
\usepackage{latexsym}
\usepackage{graphicx}
\usepackage{graphics}
\usepackage{pstricks}

\begin{document}

\newcommand{\er}{\end{eqnarray}}
\newcommand{\br}{\begin{eqnarray}}
\newcommand{\be}{\begin{equation}}
\newcommand{\ee}{\end{equation}}
\newcommand{\epe}{\end{equation}}
\newcommand{\bea}{\begin{eqnarray}}
\newcommand{\eea}{\end{eqnarray}}
\newcommand{\ba}{\begin{eqnarray}}
\newcommand{\ea}{\end{eqnarray}}
\newcommand{\epa}{\end{eqnarray}}
\newcommand{\ar}{\rightarrow}
\newcommand{\oh}{\displaystyle{\frac{1}{2}}}
\def\I{{\cal I}}
\def\A{{\cal A}}
\def\F{{\cal F}}
\def\a{\alpha}
\def\b{\beta}
\def\r{\rho}
\def\D{\Delta}
\def\R{I\!\!R}
\def\l{\lambda}
\def\d{\delta}
\def\T{\tilde{T}}
\def\k{\kappa}
\def\t{\tau}
\def\f{\phi}
\def\p{\psi}
\def\z{\zeta}
\def\G{\Gamma}
\def\ep{\epsilon}
\def\hx{\widehat{\xi}}
\def\na{\nabla}
\newcommand{\Vslash}{V\!\!\!\!/}
\newcommand{\Dslash}{\partial\!\!\!\!/}
\newcommand{\zslash}{\zeta\!\!\!\!/}
\newcommand{\mus}{\mu\!\!\!\!/}

\begin{center}

{\large The Doublet Extension of Tensor Gauge Potentials and a Reassessment of the Non-Abelian Topological Mass Mechanism}


\vspace*{1.2cm}

{\large M. Botta Cantcheff $^{\dag, \ddag}$
\footnote{e-mail: botta@fisica.unlp.edu.ar}, J. A. Helayel-Neto,  $^{\ddag}$\footnote{e-mail:
helayel@cbpf.br}}

\vspace{.2 in}

$^{\dag}${\it IFLP-CONICET and Departamento de F\'{\i}sica\\ Facultad de Ciencias Exactas, Universidad Nacional de La Plata\\
CC 67, 1900,  La Plata, Argentina;}
\vspace{.1in}

$^{\ddag}$ \emph{Centro Brasileiro de Pesquisas F\'{\i}sicas (CBPF),\\
Rua Dr. Xavier Sigaud 150, 22290-180, Rio de Janeiro, RJ, Brasil.}

\vspace{3mm}

\end{center}

\begin{abstract}
\noindent

A well-defined local non-Abelian gauge connection involving a rank-$p$ gauge $B$-field was introduced a decade ago. This was achieved by introducing doublet groups and doublet-assembled connections that can act on a doublet of matter fields, and so, standard gauging procedures and minimal substitutions are applicable.

 We provide here a summarized version of this formalism. We also build up doublet-extended gauge-invariant actions for bosonic and fermionic matter fields, and discuss the appearance of novel topological quantities in these doublet-type gauge models. A partner action for higher-spin fields appears in the doublet version of the fermionic matter sector. As an application of the proposed formalism, a gauge-invariant Chern-Simons action in four dimensions is built up; a second order quadratic Yang-Mills type action  can also be defined in this doublet framework, but we show that they can only be invariant under a Lie subgroup of our extended gauge symmetry. We finally carry out the task of studying the topological mass generation mechanism in the simplest model combining these actions into a non-Abelian generalization of the Cremmer-Scherk-Kalb-Ramond theory that we also show to be power-counting renormalizable.
\end{abstract}

\section{Introduction}

The (Abelian) Kalb-Ramond field \cite{kr0,kr} (KR), $B_{\mu\nu}$,
is a two-form field which appears in the low-energy limit of
String Theory \cite{it5}, in Quantum Gravity and Supergravity \cite{it6} and in
several other frameworks in Particle Physics \cite{aplic}. In
particular, attempts to assign mass to gauge field
models in four dimensions take into account this sort of field accompanied by a
one-form gauge field \cite{tm0,tm,la}, in order to re-create the topological mass mechanism of three dimensions.
In this sense, it becomes widely accepted that the electroweak breaking and the mass generation for the matter should be originated from a more fundamental description. The topological mass mechanism sets out to address this question: the mass for the gauge boson is not generated by means of a breaking; we rather have extra physical degrees of freedom inherited from a more fundamental level. We attempt here to deepen the investigation of properties of the 2-form gauge potential, by trying to better understand its non-Abelianisation, the possibility to realise its non-minimal coupling to matter and to work out its behaviour in the ultraviolet regime and quantum consistency.


The Abelian symmetry of the KR field is similar to that of a
$1$-form gauge field \cite{la}:
\begin{equation}
B_{\mu\nu} \to B_{\mu\nu} + \frac 12 (\partial_{\mu}\beta_{\nu} - \partial_{\nu}\beta_{\mu}) ,
\label{trKR}
\end{equation}
where $\beta_\nu$ is a $1$-form parameter.
 The old question is how
to associate the parameter $\beta_{\mu}$ to the manifold of
a general Lie group \cite{otro,ultimo}?
From the physical point of view, it is essential to ask whether a {\it
genuine} gauge theory may be formulated for this field, {\it i.e},
if the two-form gauge potential may be stated as a connection on
some group manifold. This is important because, as it is well-known,
this structure would be crucial for the identities which determine the renormalisability
or the eventual finiteness of physical models. In particular, in ref.
\cite{h}, it was argued that massive (non-Abelian) gauge models
\cite{tm0,tm} based on a gauge KR field, $B_{\mu\nu}$,
and a usual one-form, $A_{\mu}$, are ill-defined in four space-time
dimensions. The first task of this article is to clarify
the group structure underneath non-Abelian KR fields and to show that non-Abelian
theories based on them may be formulated in a similar way to the Yang-Mills-Chern-Simons
theories in $2+1$ dimensions, which are known to be finite
\cite{helpi}. We however accomodate the KR in an extended scenario, which we shall refer
to as the Doublet Formalism (DF).

Another crucial question that arises is how to define a minimal coupling of
this field with matter fields, with the interaction with gauge
fields appearing by replacing partial derivatives of the matter
fields by covariant ones in the free Lagrangian. This is directly
related to the charge conservation laws via Noether's theorem. To
do this, local gauge transformations for matter fields need to be
defined. Up to now, this is unknown for transformations which
involve a $1$-form parameter. Some alternatives for these
questions were proposed \cite{0it}, where its expected applications in
gravitation with torsion and $p$-form cosmology are emphasized
\cite{0it,it11,it12,it13, nunes-3-tensor}.

The problem was discussed first considering possible representations through
singlet tensor/spinor spaces, such that the KR field
could be built in the connection \cite {dkr,krvec,gkr}. However, in ref. \cite{krvec}
 (also in \cite{teit})
it has been shown that many
difficulties involving Lorentz invariance of physical models appear
in the non-Abelian case, and furthermore, the extension to non-flat space-times seems to be hard.
A mathematical framework where these difficulties are solved was finally proposed in Ref. \cite{DF} some years ago. According to this,
one may have a well defined two form gauge field and tackle all these problems, the basic proposal to achieve this is to relax the
conventional representations of groups, imposed in the preceding approaches. This allows us to
construct well defined gauge models for KR fields, which may be
minimally coupled with matter fields in a natural way. Once more,
the simplest solution of the problem arises from considering
doublets of tensors of different ranks as a representation for a
Lie group. This kind of idea has been successfully used before, to solve
other algebraical issues related to dualities and the Hodge map in field theory \cite{dob}.
In fact, by considering a doublet field representation, we are able to
include a $1$-form parameter in an {\it exponential-like}
symmetry/transformation law: \be\label{transdob0} \d \left(
\begin{array}{c}
  \phi \\
 \phi_\mu
\end{array}\right)= \left( \begin{array}{c}
i\alpha \phi \\ i \alpha \phi_\mu + i \b_\mu
\phi
\end{array}\right)
= i \left( \begin{array}{cc}
  \a & 0 \\
  \b_\mu & \a
\end{array}\right)
\left(\begin{array}{c}
  \phi \\
  \phi_\mu
\end{array}\right)
\ee where the variation of the fields is proportional to
themselves and to the group parameters \cite{DF}. These simple expressions
solve the problem of writing this kind of transformation law in a
simple and satisfactory way, and are the key to define the group
operations involving an $1$-form parameter. Notice that, without a
doublet representation (and a scalar parameter $\alpha$),
individual fields ($\phi$ and $\phi_\mu$) can never be combined
with an $1$-form $\beta_\mu$ to give a tensor of the same type,
and to define their variations. Clearly, (\ref{transdob0}) is the most general rule
 where both $\b$ and a minimal number of matter
 fields appear linearly and in a Lorentz-invariant way, such as
 was argued in refs. \cite{krvec,teit}.
  This idea may sound technically trivial but it is meaningful, it has never been used
  before as the cornerstone for
 a gauge principle generating the two-form field.
In this approach everything can be expressed in a manifestly
covariant form, and the generalization to curved space-time becomes rather immediate.
In this paper, we describe this formalism in detail, simplifying and clarifying the language of our original proposal \cite{DF} in order to make it useful; and finally apply it to construct a well-defined non-Abelian generalization of the topologically massive CSKR theory.
Most recently, we would like to point out that, in the work of ref. \cite{matsuo-ho}, the authors present a generalization of
the Stueckelberg mechanism for a massive 2-form potential with a
suitable transformation law for the gauge variation of the corresponding
3-form field strength.

This work is organized as follows: in Section 2, we summarize some of the the main ingredients of DF: we explicitly
find out the Lie group corresponding to these transformations and build up
the covariant derivative with the generic tensor field being part
of the connection is defined;
we also define some extra tools as Hodge duality and integration in DF, and furthermore present a new topological index associated with the extension of the group. In Section 3, we define the non-Abelian Chern-Simons theory in four dimensions and show its explicit connection with the so-called BF-theories. A Yang-Mills-type quadratic action is also analyzed. The construction of gauge invariant matter lagrangians is presented in Sec 5.
Finally, in Section 6, we study the power-counting renormalizabilty of the Yang-Mills-Chern-Simons model in four dimensions and discuss the topological mass generation mechanism. Concluding Remarks are cast
in Section 7.

\section{A Summary of the the Doublet Formalism.}

Let $(M, g_{\mu \nu})$ be a general $D$-dimensional space time
and $G$ be a Lie group whose associated algebra is ${\cal G}$;
${\tau}^{a}$ are the matrices representing the generators of the
group with $a= 1,\ldots , \mbox{dim}\:G$; $\tau^{abc}$ are the
structure constants ($[\tau^a , \tau^b] = \,\tau^{abc}\,\tau^{c}$).

As mentioned, consider the general transformations
\bea  \d\phi &=&  i\alpha \phi  \nonumber\\
\d\phi_\mu &=& i \alpha \phi_\mu + i \b_\mu \phi \label{transdob}\eea
where the doublet of parameters $(\alpha , \beta)$ consists of two
Lie algebra valued $0$- and $1$-forms respectively \footnote{We
assume them in a matricial representation of the algebra.}. Let us
denote the doublet of fields by $\Phi \equiv (\phi ,\phi _{\mu
})$. Thus, this transformation may be formally expressed as \be
\delta_{\a, \b} \Phi = ( I\a + \sigma \b ) \Phi \ee where
$I,\sigma $ are $2\times 2$ matrices, satisfying the algebra:
\be\label{algebra}
I \sigma = \sigma \,\,,\,\, I^2 = I\,\,, \,\,\sigma^2 =0
\ee
From expression (\ref{transdob0}), the simplest representation of this algebra is
\bea
I &=& \left( \begin{array}{cc}
  1 & 0 \\
  0 & 1
\end{array}\right) \,,\,\\
\sigma &=&\left( \begin{array}{cc}
  0 & 0 \\
  1 & 0
\end{array}\right)\, .
\eea
Notice that there is another interesting representation by taking $\sigma$ to be a Grassman parameter.

The product of two elements of the algebra is well defined and is
naturally given by the usual matrix product \be \label{matprod}
\delta_{\a', \b'} \,\delta_{\a, \b} \Phi =  I (\a'\, \a ) + \sigma (\b'\,\a + \a' \,\b) \Phi , \ee

This is the most general associative product (such that the upper component of the matrix (\ref{transdob0}) vanishes) we can construct in this way \cite{DF}.

Thus, the algebra elements may alternatively be expressed in terms of
\emph{doublets}, ordered pairs of $0$- and $1$-forms.
The product of
doublets reads as: \be \left(  \alpha\;,\;\beta_{\mu}\right)
\left(  \phi\;,\;\phi_{\mu
}\right)  =\left(  \alpha\phi \;,\;\alpha\phi_{\mu}
+\beta_{\mu}\phi\right) \label{defprod} \ee
thus the symmetry transformation is
\[
\left(  \delta\phi\;,\;\delta\phi_{\mu}\right)  =i\left(
\alpha\;,\;\beta_{\mu}\right)  \left(  \phi\;,\;\phi_{\mu}\right)
= \left(
 i \alpha \phi\,,\,i  \alpha \phi_\mu + i  \b_\mu \phi
\right) ,
\]
and the identity represents as $(1 , 0)$.

 Using the expression (\ref{transdob}) for the infinitesimal transformation,
 it may be easily shown that a group element may be expressed as usually
 \be\label{g}
 g=\exp i \G
 \ee
 where the algebra parameter in this case writes \be\label{alg-elements} \G = ( I\a^a + \sigma \b_\mu^a )\,\tau^a .\ee
  Notice that by virtue of the property $\sigma^2=0$, expression (\ref{g}) reduces to be linear in $\b$. For instance, consider a very particular example where the Lie group has the structure $G_{(\a)}
\times G_{(\b)}$, such that the generators of the $\a$ and $\b$-sectors are different: $\G =  I\a^a \tau^a + \sigma \b_\mu^a \tau'^a\,$;  $[\a
,\b]=0$, then the group element results $ g(\a , \b_\mu ) = ( e^{i\a} , i \b_\mu
e^{i\a})$ \cite{dob}. Except this special case, the exponential form (\ref{g}) is mnemonically more natural and easier for computations and algebraic manipulations.

The set of exponentials (\ref{g}) constitutes a group since the Lie Algebra can be extended to the doublet algebra (DA) by taking the following set of generators:
\be
\{ \tau^a\, ; \,\,\tau^{a\mu}\equiv \sigma \tau^a e^\mu \}\, ,
\ee
 where $e^\mu$ denotes a basis of the space-time one-forms. These elements close the following extended algebra:
\be\label{DA}
[ \tau^{a} , \tau^{b} ] = f^{abc} \tau^{c}\,~;~~\,
 [\tau^{a\mu} , \tau^{b} ] = f^{abc} \tau^{c\mu} \,~;~~\,\,[\tau^{a\mu} , \tau^{b\nu} ] = 0~~.
\ee
So, exponents of the form (\ref{alg-elements}) can be expressed as $\G = \a^a \,\tau^{a}\, + \,\b_\mu^a\, \tau^{a\mu}$. The subset $\beta\equiv 0$ defines the conventional Lie soubgroup/subalgebra.
The rest of the construction is standard and follows in a straightforward way from what given above.

\subsection{General doublets of $(1,r+1)$-tensors as gauge connections.}

Let us consider the most general representations of doublets introduced in Ref. \cite{DF}, which consist in pairs of \emph{tensors} $(\phi^q_{\,\,p} , \phi^{q'}_{\,\,p'})\in
\Pi^q_{\,\,p} \times \Pi^{q'}_{\,\,p'}$ ($\Pi^q_{\,\,p} $ denotes the standard set of tensors of type $(q,p)$
\footnote{In other similar formalisms, only \emph{forms} and exterior calculus are considered}).
So, the symmetry transformation can be built over doublets of arbitrary
order using the same idea, in any of the spaces $(\phi^q_{\,\,p} , \phi^{q'}_{\,\,p'})\in
\Pi^q_{\,\,p} \times \Pi^{q'}_{\,\,p'} \;, \forall p , q, q', r$ which takes values in a
representation of the Lie group. For simplicity, let us take arbitrary pairs of tensors with coincident covariant type $\Phi = (\phi_p ,
\phi_{p'})$ and $(\a , \b_{r}) \in \Pi_{0} \times \Pi_{r}$ ($r \equiv p'-p$), and
the $r$-generalized connection reads as ${\cal A}=(A_{1}, B_{r+1})
\in \Pi_{1} \times \Pi_{r+1} $ and so on. In this case, in view of
(\ref{transdob0}), the symmetry can be written as below:
 \be\label{transdobk} \d
\left(
\begin{array}{c}
  \phi_{p} \\
 \phi_{p+r}
\end{array}\right)=  i \left( \begin{array}{cc}
  \a & 0 \\
  \b_r & \a
\end{array}\right)
\left(\begin{array}{c}
  \phi_{p} \\
  \phi_{p+r}
\end{array}\right)\; .
\ee

For simplicity consider now $q=0$.
We introduce the {\it partial derivative} of a $(0,p)$ tensor
$T_{p}$ as a $(0,p+1)$ tensor given by
\[
T_{p}=T_{\mu_{1}...\mu_{p}}dx^{\mu_{1}}\otimes...\otimes
dx^{\mu_{p}},
\]
as
\[
\partial T_{p}:=\partial_{\mu}T_{\mu_{1}...\mu_{p}}dx^{\mu}\otimes dx^{\mu_{1}
}\otimes...\otimes dx^{\mu_{p}}.
\]
So, we can define the partial derivative of a doublet as the
doublet consisting of the partial derivatives \be
\partial (\phi_p , \phi_{p+r})
\equiv ( (\partial \phi_{p})_{p+1} , (\partial
\phi_{p+r})_{p+r+1}).\ee It is easy to verify that this definition
is consistent with the Leibnitz rule for the product of doublets.

The tensor product of two doublets of arbitrary orders and
types,
 is the simple generalization of the rule (\ref{defprod}):$
    (A,B)(A',B') = (A \otimes A'  , A \otimes B' + B \otimes A' )$.
 Next, one can
 define the [nonexterior] covariant
derivative of a $(p, p')$-doublet, $\Phi =(\phi_p ,
\phi_{p'})$, as a $(p+1, p'+1)$-doublet: \be D \Phi =
\partial \Phi - i [{\cal A} \,, \,\Phi]= \left(
\partial \phi_p- i  [ A_1 , \phi_p ] \; , \;
\partial \phi_{p+r} - i  [ A_1 , \phi_{p+r} ] -i [ B_{r+1} , \phi_{p} ]
\right)\, ,\ee where the connection must be a doublet valued on any Lie algebra: \be\label{rdob} {\cal
A} \equiv ( A_{1} , B_{r+1}) = ( A^a_{1} , B^a_{r+1}) \,\tau^a\ee of order $r=p-p'$. These are what we wish call \emph{Doublet-Connections}, and constitutes one of the most meaningful achievement of the present approach since \emph{any-rank tensor} canonically appears in a gauge connection on an arbitrary Lie group, in a way that avoid all the old conflicts with Lorentz invariance \cite{krvec}
  \cite{teit}. We would like to highlight that the doublet of eq. (\ref{rdob}) sets out as a genuine non-Abelian connection, and the gauge transformations below, eqs. (\ref{A}) and (\ref{B}), show that an Abelianization of the problem is actually not possible for generic fields and arbitrary transformations $\a, \b)$, except in the separable case $[\a ,\b]$=0.


 Imposing that $ g D \Phi = D'
\Phi' $, and using $ D' =
\partial - i [{\cal A}'\,, \;]$ and $\Phi'_{} =g \Phi_{}$, we obtain the
transformation law for the connection:\be\label{tr-gauge} {\cal A}' = g(\a ,\b)
{\cal A} g(-\a ,-\b) -i( \partial g(\a ,\b)  ) g(-\a ,-\b), \ee
whose infinitesimal expression is \be\label{tr-gauge-inf}\d {\cal A} =\partial (\a ,\b)
- i[{\cal A},(\a ,\b)]= D (\a ,\b)\ee (The canonical
curvature tensor ${\cal F} \in \Pi_{2} \times \Pi_{2+r} $
transforms as ${\cal F}' = g(\a ,\b){\cal F}g(-\a ,-\b) $.) ;
which reads, in terms of the doublet components: \be \label{A} \d
A =
\partial \a - i [ A, \a ] \; , \ee \be \label{B} \d B = \partial
\b -i [ B, \a ] -i [ A, \b ]\; . \ee

So for instance, in terms of tensor components, the curvature tensor ${\cal F}
= ( F_2 , h_3 ) \in \Pi_{2} \times \Pi_{3} $ results as \be
\label{F} F_{\mu \nu} = 2
\partial_{[\mu} A_{\nu]} + i [ A_{\mu} , A_{\nu} ] \ee \be
\label{h} G_{\mu \nu \r} = 2\partial_{[\mu} B_{\nu] \r} + i (
\left[ B_{\mu  \r} ,\, A_{\nu} \right] + \left[A_{\mu} , B_{\nu\r}\right] )\;
\ee
which, expressed in terms of the algebra generators reads
\bea
\label{h-indices}G_{\mu \nu} = G^c_{\mu \nu \r} \tau^{c \rho} &=& 2\partial_{[\mu} B^c_{\nu] \r}\, \tau^{c \rho} + i (
B^a_{[\mu | \r} \, A^b_{\nu]} +
 A^a_{[\mu} B^b_{\nu]\r} )\,\, f^{abc} \tau^{c \rho}\;=\nonumber\\
 &=& 2\partial_{[\mu} B^c_{\nu] \r}\,\tau^{c \rho} + 2i\;
 B^a_{[\mu | \r} \, A^b_{\nu]}\,\, f^{abc} \tau^{c \rho} \;
\eea
where brackets $[\,,]$ stand for anti-symmetrization over the indices $\mu,\nu$, and
the symbol $|$ before the $\r$ index means that $\r$ is not to be
anti-symmetrized.

Let us remark that DF  \cite{DF} is more general than often definitions of the $B$-field and its associate curvature $G$:
recall it describes connections $(A_1, B_p)$, where $B$ may be general contravariant tensors (or spinors \cite{nogue}) rather only $p$-forms.

This allows us to describe all tensor (and spinor) types in a connection on an arbitrary group and generalize the gauging principles, with tensors as gauge parameters. In the general formula (\ref{h}), $B$ and $G$ are not totally anti-symmetric and contain more components, then, the so-called Kalb-Ramond gauge field must be identified with its totally anti-symmetric part: $
B_{\mu\nu}\equiv  B_{[\mu\nu]}$, and the totally antisymmetric part of the curvature doublet is the usual $(2,3)$-form pair: $(F, H) \equiv (F_{[\mu\nu]}, H_{[\mu\nu\rho]})$, where, according to (\ref{h})
\be
\label{H} H_{\mu \nu \r} = 2\partial_{[\mu} B_{\nu\r]} + i \left([
B_{[\mu  \r} \,, \, A_{\nu]} ] + [ A_{[\mu}\, , B_{\nu\r]} ]\right)\; \equiv d \wedge B + A \wedge B + B\wedge A . \ee

 So the Kalb Ramond field may be found as a component of the exterior connection in the exterior covariant derivative, defined as the totally anti-symmetric part of the covariant derivative defined above (where tensor products $\otimes$ must be substituted by exterior ones $\wedge$). For arbitrary rank:
\be\label{D-wedge}  D \wedge \Phi \equiv
\partial \wedge \Phi - i {\cal A} \wedge \Phi= \left(
d\, \phi_p- i  A_1 \wedge \phi_p \; , \;
d\, \phi_{p+r} - i  A_1 \wedge \phi_{p+r} -i B_{r+1}\wedge \phi_{p}
\right)\, ,\ee
and the doublet field strength is the curvature of this derivative:
\be\label{Curvatura-r}
{\cal F} \equiv (F_2 , H_{r+2}) = (\,\, d \wedge A_{1} + A_1 \wedge A_{1} ,\,\,d \wedge B_{r+1} + A_1 \wedge B_{r+1} + B_{r+1}\wedge A_1).
\ee
This satisfy the Bianchi identities:
\be\label{bianchi}
D \wedge {\cal F} \,=0\,.
\ee
Using (\ref{D-wedge}) this splits as:
\be
(d\, - i  A) \wedge\,F =0  \,\,\,\,\,\, \,\,,\,\, \,\,\,\,\,\,\,\,\,\,(d\, - i  A) \wedge H -i B\wedge\,F\,\,=0\,.
\ee
These relations are a consequence of the Jacobi identities, which are closely the associative property of the algebra. We should recall however, that DF admits natural generalizations to \emph{non-associative} algebras through the introduction of a metric in the base manifold (more details may be found at the original article \cite{DF}).

We have noticed that
this general DF is coincident with other formalisms whenever one adopts specific choices; e.g.,
whenever particularized to exterior calculus for pairs of $p,p+1$-rank forms, it is similar to the so-called generalized exterior calculus developed in \cite{robinson}; and furthermore, for the simplest doublet connection $(A_1, B_2)$, our formalism coincides with the more recent construction of \cite{sav}.

\subsection{Hodge map and integration in the DF.}

The formalism above requires some extra structure in order to define topological actions and gauge theories in general. Concepts as the Hodge operation and integration on manifolds shall also be extended according to this formalism in a consistent way.

The problem of defining self(/anti-self)-duality relations for arbitrary forms, space time dimension and signature, may be solved by introducing doublets of this type and generalizing the Hodge operation accordingly, which allowed to avoid the conflict with the signal of the contraction of two Levi-Civita tensors \cite{dob}.

Let us consider a $d$-dimensional space-time with signature, $s$, and a generic
forms doublet $\Phi \equiv (\phi_p\,,\,\phi_q) $ in the space $\Lambda_{p,q}\equiv \Lambda_p \times \Lambda_{q}$.
Thus, one may define a {\it Hodge-type} operation for these
 objects
 by means of
\be\label{hodge}
\mbox{}^{*} \Phi \equiv (  \,S_q \mbox{}^{*} \phi_q  \, , \,S_p \,\mbox{}^{*} \phi_p ),
\ee
where $|S_p| = 1$ is properly defined in order to get consistency with:
\be\label{nilpo}
\mbox{}^{*} (\mbox{}^{*}\Phi ) = \Phi .
\ee
But the double dual operation, for a generic $p$-form $A$ depends on the signature ($s$) and dimension of the space-time in the form:
$\mbox{}^{*}(\mbox{}^{*}A) = (-1)^{s+p[d-p]} \, A \,$. Therefore, (\ref{hodge}) and (\ref{nilpo}) imply:
\be\label{Simp}
(-1)^{s+p[d-p]} S_{d-p} S_p   = 1 .
\ee
 $\mbox{}^{*}$ applied to doublets is defined such that
 its components are interchanged
 with a supplementary factor (\ref{hodge}) satisfying (\ref{Simp}), which is fulfilled if
 \be\label{S}
S_{d-p} S_p   = (-1)^{s+p[d-p]}
\ee is chosen. Notice that the requirement (\ref{nilpo}) is inconsistent with the case $p=d/2$, for Lorentzian signature $s=1$. So here, we consider the generic case $p \neq d/2$ for simplicity.
  Below, we will show that  $S_p$ is to be finally fixed through physicalness requirements on the actions.

Notice that there is a well defined notion of self (anti-self)-duality for doublets,
since
 \be
 \mbox{}^{*} \Phi =\pm\,  \Phi \,
\ee
is consistent with the requirement (\ref{nilpo}). This operation may be expressed schematically
in terms of the algebra elements ($I,\sigma$) as:
\be\label{hodge-alg}
\mbox{}^{*} I \sim \,\sigma \, \,\,\,\,\,\nonumber\\
\mbox{}^{*} \sigma \sim \,I\, \, ,
\ee up to signal factors $S_{p,q}$ that depend on the tensor coefficients.

\subsection{Integration}

The present formalism admits more general definitions of integration that might be useful in certain context related to topological objects (we are going to discuss a bit this subject in the following section) \cite{next}. In fact we naturally can define a \emph{pair} of integrals of $(A_{p} , B_{d})$ for a \emph{doublet} of manifolds $(\Sigma_p, M)$, provided that $\Sigma_p$ is a $p$-dimensional \emph{submanifold} of $M$,
 \be\label{int-pair}
{\cal I} \equiv \int_{(\Sigma_p, M)} (A_{p} , B_{d})\, \equiv \,\left(\,\,\int_{\Sigma_p} A_{p} \,\,\,\,\,\,\,\,\,,\,\,\,\,\,\int_{M} B_{d} \,\,\right)
\ee
 In order to define actions, one may combine the components of ${\cal I}\equiv(I_p,I_d) $ to define one number as the integral of $(A_{p} , B_{d})$ over $(\Sigma_p, M)$ in several ways. The natural meaningful prescription is the sum
 \be\label{int-gen}
I\equiv I_p + I_d= \int_{(\Sigma_p, M)} (A_{p} , B_{d})\, \equiv \,\,\int_{\Sigma_p} A_{p} \,\,+\,\,\,\int_{M} B_{d} \,\,\,\,.
\ee
 In contexts where there is not a natural submanifold $\Sigma_p$, the only geometric (relativistic invariant) definition, reduces just to the second term:
\be\label{int}
I\equiv \int_{ M} (A_{p} , B_{d})\, = I_d =  \,\,\int_{M} B_{d} .
\ee

In the particular case $p=d-1$, there is a natural (relativistic) hypersurface $\Sigma_p = \partial M$, then one can define the integral over $M$ by means of:
\be\label{int-d}
\int_M (A_{d-1} , B_{d})\, \equiv \,\,\int_{\partial M} A_{d-1} \,\,+\,\,\,\int_{M} B_{d} \, .
\ee
Even though in the context of topological configurations and invariants the boundary term may become important,
this integration allows us to define actions. In such a context or in a boundary-less manifold, the relevant part of the integral reduces to the second term. This recipe is in line with representing $\sigma$ as a grassmannian variable, since the integral coincides with taking a derivative with respect to it.

\subsection{New Topological Charges}

As an application, let us consider a fiber bundle with a $(d\geq 4)$-dimensional base manifold $M_{d}$, whose principal bundle is given by the action of an arbitrary Lie Group $G$. Consider in addition, a four-dimensional submanifold $M_4$ embedded into $M_d$. So in DF, one may define a generalized doublet gauge connection ${\cal A}=(A,B)\equiv A_\mu^a \,\tau^{a}\, + \, B_{\mu \nu_1 \nu_2 \dots}^a\, \tau^{a\nu_1 \nu_2 \dots}$, where here $B$ is a totally anti-symmetric $(d-2)$-form, i.e. one must set $\tau^{a\nu_1 \nu_2 \dots}\equiv \sigma \tau^a e^{\nu_1} \wedge e^{ \nu_2} \dots $.

 According to the definition (\ref{int-pair}), the gauge invariant integral
\be\label{topologic-number}
{\cal I}\equiv \int Tr {\cal F}\wedge {\cal F} = \left( \int_{M_4}Tr\, (F\wedge F)\,\,, \,\,\int_{M_d}Tr\,(F\wedge H \,+ \,H \wedge F)\right)=(I_4 , I_d)
\ee
is a meaningful doublet of topological quantities. The first component is the \emph{instanton} number or Chern-Pontryagin index in four dimensions, but the second one is a new one. Here we have used the trace in the ordinary Lie group, such that $ Tr (\tau^{a}\tau^{b}) = \delta^{ab}$, which is related to the inner product in the Lie algebra. This aspect will be clarified in the next Section.

 Notice that this quantity is defined on a $d$-dimensional manifold but clearly, although independent, it is closely related to the number of instantons in the $M_4$-projected Yang Mills theory \footnote{We are tempted to suggest a relation of duality between them.} and it shall encode relevant topological information. In a forthcoming work, we shall study topological configurations where this number can be physically interpreted \cite{next}.

  Then, let us observe that the doublet of currents (\ref{topologic-number}) is conserved. In fact, the related topological current density may be expressed as:
 \be\label{J}
J\equiv {}^{*}{\cal L} = (S_3 {}^{*}L_3 , S_{d-1} {}^{*}L_{d-1})
  \ee
 where
\be\label{Lcs}
{\cal L} \equiv (L_3 , L_4) = \left( A d A - \frac{2i}{3} A \wedge  A \wedge A \,\,, \,\, A \wedge d {
B} + {B} \wedge d  A  -  i 2 A \wedge A\wedge
{ B}  \right) \, ,
\ee which may be written in terms of the doublet connection in the suggestive form:
\be
{\cal L} \,=\; Tr \; \left( {\cal A} d {\cal A} - \frac{2i}{3} {\cal A} \wedge {\cal A} \wedge {\cal A}
\right).\ee

  The exterior derivative of this doublet is precisely the integrand of (\ref{topologic-number})
   \be
{\cal I} = d{\cal L}
\ee
which, by virtue of the Bianchi identities, shows that the total divergence of (\ref{J}) vanishes.


\section{The Chern-Simons theory and non-Abelian BF models.}

The Chern-Simons action in $d$ dimensions may be rigorously defined in DF, and notably, the relation with the paradigmatic three-dimensional case is manifest \cite{DF}.

Consider the connection ${\cal A}= (A , B)$ on an arbitrary Lie group, we can define the integral doublet (\ref{int-pair}):

\be {\cal I}_{CS} [{\cal A}] \equiv
\;  \;\int\; Tr\;\left( {\cal A} d {\cal A} - \frac{2i}{3} {\cal A} \wedge {\cal A} \wedge {\cal A}
\right).\ee
The integrand in this expression is in fact the doublet of $(3,d)$-form given explicitly by Eq. (\ref{Lcs}).
Notice that the first component corresponds to the canonical Chern-Simons theory for a one form gauge field $A$ in three dimensions, while as is manifest, the doublet as a whole is what we shall identify as its generalization to  arbitrary $d$ dimensions in DF.

 Finally, we take the integration (\ref{int}) in order to define the $d$-dimensional Chern-Simons action:
\be {\cal S}_{CS} [{\cal A}] \equiv -k
\;\int\; Tr \; \left( [ A \wedge d {
B} + { B} \wedge d  A ] -  i 2 [A \wedge A\wedge
{ B}] \right)\, .\label{CS}\ee
where $k$ denotes the inverse of the coupling constant.
This is a well defined gauge invariant topological theory, and may be recognized as the generalization of
a so-called BF theory for an arbitrary non-Abelian Lie group. The second term in the R.H.S. is not obvious and it is necessary to get the gauge invariance (\ref{tr-gauge-inf}). It is indeed straightforward to check out that ${\cal S}_{CS}$ is
gauge invariant (up to a total derivative) as
expected.

The equations of motion read:
\be\label{CSeqs}
{\cal F} = d {\cal A} - i {\cal A} \wedge {\cal A} = (\,F \,,\, H\, )=0 \,\,.
\ee

\vspace{0.5cm}

 $B \wedge F $ theories are similar to
 Chern-Simons in three dimensions and they are often associated by analogy \cite{dob}, however,
the actual connection between both never has been clearly established
so far \cite{DF}.
 In the present framework
this is indeed defined as {\it a genuine} Chern-Simons theory for a doublet connection.

In the particularly meaningful $d=4$ case, one can define the invariant integral as (\ref{int-d}), and the Chern Simons adopts the more general form:
\be {\cal S}_{CS} [{\cal A}] \equiv \;\int_{\partial M_4}\; Tr \; \left( A d A - \frac{2i}{3} A \wedge  A \wedge A \,\,\right)\,+
\;\int_{M_4}\; Tr \; \left( [ A \wedge d {
B} + { B} \wedge d  A ] -  i 2 [A \wedge A\wedge
{ B}] \right)\, ,\label{CS4}\ee
which includes a boundary term that precisely is a Chern Simons theory for the gauge field $A_\mu$ \footnote{For a boundaryless $M_4$, or asymptotic decay of $A$, this action agrees with (\ref{CS}) for dimension $d$.}. Therefore, BF theory (rigorously generalized here to non-Abelian groups) and Chern Simons theory, are manifestly part of a \emph{same} structure in DF.

\vspace{0.7cm}

\textbf{A note on the inner product}

\vspace{0.7cm}

Notice that to define the actions above we have implicitly used the trace in the ordinary Lie group, such that $ Tr (\tau^{a}\tau^{b}) = \delta^{ab}$; however, in order to define gauge invariant numbers, the necessary structure is an inner product $\langle \, ; \,\rangle$ in the extended algebra. In an ordinary Lie Algebra one has standard matricial representations and then it reduces to the ordinary trace $\langle A\, ; B\,\rangle = (Tr A B )$, but there is no an obvious way to extend this to DA. Then, if we naively define it by using the standard trace, we have that the inner product extends as below
\be\label{traceDA}
Tr(\tau^{a} , \tau^{b}) = \delta^{ab} \,~;~~\, Tr(\tau^{a\mu} , \tau^{b})= \sigma \,\delta^{ab} e^{\mu} \,~;~~\,\,Tr(\tau^{a\mu} , \tau^{b\nu})=  0~~
\ee
so that the basis of the algebra  (\ref{DA}) is not orthogonal with respect to it. In the present work we shall consider this extension, but as we warn below, in certain specific cases one may run into troubles in defining invariant actions.

\section{The Yang-Mills action}

Next, we may canonically define Yang Mills theories for doublet connections in $d$ dimensions ${\cal A} = (A_{1}\,\,, \,\,B_{r+1})$, $r \leq d-2 $.
The field strength tensor is ${\cal F}=(F_{2}\,,\,H_{r+2})$, and by virtue of \ref{hodge}, the curvature dual is:
\be
{}^{*}{\cal F} = (S_{r+2}\,\,{}^{*}H \,\,, \,\,S_{2}\,{}^{*}F)\, ,
\ee
which is a doublet $(d-r-2\,\,,\,\,d-2)$-form. Therefore,
\be
{\cal F}\wedge {}^{\star}{\cal F} = ( S_{r+2} F\wedge{}^{*}H \,\,, \,\,S_{2}\,F\wedge{}^{*}F \,+ \,S_{r+2}\,{}^{*}H \wedge H)
\ee
 is a $(d-r\,, \, d)$-form. Therefore, the pure Yang-Mills action can be defined as the integral (\ref{int}) of this expression, which is gauge invariant for construction:
\be\label{YM}
{\cal S}_{YM}\equiv \int Tr\,{\cal F}\wedge {}^{*}{\cal F} = \,\int\,Tr\,\left(S_{2}\,\,F\wedge{}^{*}F \,+ \,S_{r+2}\,\,{}^{*}H \wedge H\right)\;.
\ee
This may be written also as:
\be
{\cal S}_{YM}\equiv \int dx^d \,\,Tr\,\,\left( S_{d-2}S_{2}\,\,F\cdot F \,+ \,S_{d-r-2}S_{r+2}\,\,H \cdot H\right)\;,
\ee
where the dot represents the ordered contraction over all the space-time indices: $P \cdot Q \equiv P_{\mu_1\dots\mu_n} Q^{\mu_1\dots\mu_n}$.
Using (\ref{S}), we have:
\be\label{YM-volume}
{\cal S}_{YM}\equiv \int dx^d \,\,Tr\,\,\left( (-1)^{s+2(d-2)}\,\,F\cdot F \,+ \,(-1)^{s+(r+2)(d-r-2)}\,\,H \cdot H\right)\;.
\ee
The equations of motion (source-free) read:
\be\label{eqYM}
D\wedge \mbox{}^{*} {\cal F} = 0 ,
\ee
in components,
\be\label{eqYM-comp}
\left( \, (d\, - i  A_1), -i B_{r+1} \right) \wedge \left( S_{r+2}\,(\mbox{}^{*} H)_{d-r-2} \,\, ,\,\, S_{2}\,(\mbox{}^{*} F)_{d-2}\, \right)\,=0\,,
\ee
where $\mbox{}^{*}$ is the standard contraction by the Levi-Civita tensor. Using (\ref{D-wedge}) this splits in two equations:
\be
(d\, - i  A_1) \wedge\,\mbox{}^{*}H =0  \,\, ,\,\,\,\,\,S_{2}\, (d\, - i  A_1) \wedge \mbox{}^{*}F -i S_{r+2}\,\,B_{r+1}\wedge\,\mbox{}^{*} H\,\,=0\,.
\ee
Contracting these equations by the totally anti-symmetric Levi-Civita tensor, they adopt the more familiar divergence type form \footnote{In this notation the scalar product ``$ \cdot $'' also denotes the product in the Lie algebra given by the commutator, i.e: $ A\cdot B \equiv [A_\mu , B^\mu ]$.}:
\be
(\partial \, - i  A_1) \cdot H =0  \,\, ,\,\,\,\,\,\, (\partial\, - i  A_1) \cdot F -i \,(-1)^{r(d-r)}\,B_{r+1}\cdot H\,\,=0\,.
\ee


Unfortunately, this action is not invariant under the full set of gauge transformations (\ref{g}). The explicit violation of the $\b$-invariance can be computed as
\be\label{YMvariation}
\delta {\cal S}_{YM} \propto \int dx^d \,\,Tr\,\,\left( \,[\beta, F] \wedge \mbox{}^{*} H\right)\;\,\,\,\,\,.
\ee
The reason why this is not $\b$-invariant is not clear for us, but this could be due to the structure extra to the group or algebra extensions defined to construct scalars. To our mind, this can be a consequence of the ill-defined inner product in the extended doublet algebra observed above.

Nevertheless, notice that YM action \emph{is invariant} under the subgroup of $\a$-transformations (defined by fixing the parameter $\beta_\mu=0$), that is the conventional Lie group in standard gauge theory. In the model with topological mass generation studied below, we will take advantage of this fact.

\section{Matter Fields}

~

\textbf{Bosonic matter:}

~

The Hodge operation and integration are the need ingredients to define general global (and local)
 gauge invariant actions in a exterior calculus language. The standard free action for a complex scalar (a $0$-form) field $\phi$ in the adjoint representation of a group $G$ expresses as
 \be\label{scalar}
{\cal S}_{scalar}(\phi)\equiv \frac{1}{2} Tr\, \int\,d\overline{\phi}\wedge \mbox{}^{*} d\phi + \frac{m^2}{2}\int Tr\, \overline{\phi}\wedge \mbox{}^{*} \phi \, .
\ee
 In this language, additional interacting terms may be written as:
 \be\label{scalar-interaction}
 {\cal S}_{int}\equiv\sum_{n}\frac{\lambda_n}{n!} Tr\, \int\,\left(\overline{\phi}\wedge \mbox{}^{*} \phi\right)^n .
\ee
The rule to form gauge invariant action for bosonic matter fields in a four-dimensional Lorentzian space time, consists in doing the substitution:
\be\label{extension} \phi \,\, \to\,\,\Phi =  I\phi + \sigma \phi_\mu\, =( I\phi^a + \sigma \phi_\mu^a )\,\tau^a .\ee
Plugging this into the action (\ref{scalar}), using the algebra (\ref{algebra}), (\ref{hodge}), and the integration defined above, we obtain:
\be\label{doublet-scalar}
{\cal S}(\phi, \phi_\mu) = \frac{(-1)}{2} Tr\, \int\,d^4x \left(\partial_\mu\overline{\phi}\partial^\mu\phi - \partial_{[\mu}\overline{\phi}_{\nu]}\partial^{[\mu}\phi^{\nu]} + \frac{m^2}{2}\, (\overline{\phi}\phi+\overline{\phi_\mu}\phi^\mu)\right).
\ee

For $p$-form matter fields the rule is the same, the action (\ref{scalar}) is well defined for
 any type of the singlet $\phi\in\Lambda_{\,\,p}$; then, one shall extend this to a doublet as (\ref{extension})
 $\phi\to \Phi= (\phi, \phi_r)\in \Lambda_{\,\,p}\otimes\Lambda_{\,\,p+r}$, with $p,r$ arbitrary.

\be\label{extension-r} \phi \,\, \to\,\,\Phi \equiv  I\phi + \sigma \phi_r\, =( I\phi^{a}_{\mu_1....\mu_p} +
 \sigma \phi^{a}_{\mu_1....\mu_{p+r}} )\,\tau^a .\ee

Then the free action in $d$ dimensions becomes
\bea\label{doublet-form-matter}
{\cal S}(\Phi) = \frac{1}{2} Tr\, \int\,d^dx  \left(S_{d-p-1} S_{p+1}\,\,\partial_{[\mu}\overline{\phi}_{\mu_1...\mu_{p}]}\partial^{[\mu}\phi^{\mu_1...\mu_{p}]} \right.+\nonumber\\
+ S_{d-p-r-1} S_{p+r+1}\,\, \partial_{[\mu}\overline{\phi}_{\mu_1...\mu_{p+r}]}\partial^{[\mu}\phi^{\mu_1...\mu_{p+r}]}
  +\nonumber\\ \left.\frac{m^2}{2}\,(\,\,(S_{d-p} S_{p})\,\,\overline{\phi}_{[\mu_1...\mu_{p}]}\phi^{[\mu_1...\mu_{p}]}+\,\,(S_{d-p-r} S_{p+r})\,\,\overline{\phi}_{[\mu_1...\mu_{p+r}]}\phi^{[\mu_1....\mu_{p+r}]})\right).
\eea where the signal in front of each term is determined from the formula (\ref{S}).

Locally invariant actions are obtained from this one by minimal substitution of the covariant derivative $d\to D$
that includes standard gauge one form $A_1$ and the new tensor field $B_{1+r}$.

~

\textbf{Fermionic matter:}

~

By performing the formal replacement $p=1/2$ in the rule (\ref{extension-r} ) above, we obtain the doublet of fields we have to consider to construct gauge invariant actions that include Dirac fields in the matter sector. So then,
\be\label{ferm-extension-r} \psi_{(1/2)} \,\, \to\,\,\Psi \equiv  I\psi_{(1/2)} + \sigma \, \zeta_{(1/2) +r}\, =( I\psi^{a}_{} +
 \sigma \zeta^{a}_{\mu_1....\mu_{r}} )\,\tau^a .\ee
   Particularizing now for the simplest $d=4$ case, $\psi\equiv\psi_{1/2}$ is a spinor field, while its partner $\zeta_{(1/2) + 1}$ carries spin $3/2$ which behaves as a (massive) Rarita-Shwinger-type field, as it shall become clear in what follows below.

The global invariance under doublet gauge transformations is manifest whenever we write the Dirac action in the language of forms:

\be\label{dirac}
{\cal S}_{Dirac}\equiv \int Tr\,\,\,\bar{\psi_{}} \, \,\gamma\, \wedge {}^{*} (d \psi_{}) \, + \, i m \int Tr\, \overline{\psi}\wedge \mbox{}^{*} \psi,
\ee
where the gamma matrices are thought as components of a one-form (whose components are \emph{matrices}) $\gamma\equiv\gamma_\mu$, in order to simplify the notation; the spinor indices are all contracted in the standard way, and the trace contracts the group internal indices, and finally, the Hodge operation $*$ acts only on the tensor indices $\mu, \nu,...$.

In the mass term the product $\wedge$ does not play any role while we do not substitute $\psi$ by the doublet $\Psi$. Then plugging (\ref{ferm-extension-r}) into this action we obtain the globally invariant fermionic action:
\be\label{DF-dirac-doublet}
{\cal S}_{f}(\Psi) \equiv \int Tr\,\,\,\bar{\Psi_{}} \, \,\gamma\, \wedge {}^{\star} (d \Psi_{}) \, + \, i m \int Tr\, \overline{\Psi}\wedge \mbox{}^{\star} \Psi ,
\ee
which, by using the DF algebra and integration, reads
 \be\label{DF-dirac}
{\cal S}_{f}(\Psi) = {\cal S}_{Dirac} (\psi) +  \int Tr\,\,\,\bar{\zeta} \, \,\gamma\, \wedge {}^{\star} (d \zeta) \, + \, i m \int Tr\, \overline{\zeta}\wedge \mbox{}^{\star} \zeta,
\ee
where $\zeta$ has to be thought as a one form, while the spinor index is implicit (and contracted with $\gamma$'s as standard). This sector is new and may be written in components as
\be\label{D-RS}
{\cal S}(\zeta) = \int Tr\,\,\,\bar{\zeta}_{[\mu } \, \, \,\gamma_{\nu]}\,\,\, (\partial^\mu \zeta^\nu) \, dx^4 \,+ \, i m \int Tr\, \overline{\zeta}_\mu \zeta^\mu ,
\ee
This action gives dynamics to the spin-$3/2$ field. Let us now understand how close it is related to the Rarita-Schwinger field and how the spin-$1/2$ mode propagates. The equation of motion is
\be\label{eq-D-RS}
\gamma^\mu \, \partial_{[\mu } \, \,\zeta_{\nu]}\, = \, i m \zeta_\nu ,
\ee
which looks like as a ``dual" to a conventional Rarita-Schwinger field. Notice that, by choosing the Rarita-Schwinger gauge: $\gamma^\mu \, \,\zeta_\mu =0$, this equation of motion reduces to the Dirac equation for the component fields:
\be\label{eq-D}
\Dslash \, \,\zeta_{\nu}\, = \, i m \zeta_\nu \,\,,
\ee
which describes the propagating modes.
In fact, Eq. (\ref{eq-D-RS}) defines a way of generalizing the Dirac equation ($r=0$), to any $(1/2) + r$-form $\zeta$:
\be\label{eq-ferm}
\gamma^\mu\,(\partial_{[\mu}\, \zeta_{\mu_1....\mu_r]})\,=\, i m \zeta_{[\mu_1....\mu_r]} \,\,.
\ee
By imposing the Rarita-Schwinger-like gauge as
\be \gamma^{\mu_i} \, \zeta_{\mu_1\dots \mu_i \dots\mu_r } =0 ~~~~~~ \forall i=1,\dots,r\,\,,\ee eq. (\ref{eq-ferm}) reduces to:
\be
\gamma^\mu\,\partial_{\mu}\, \, \zeta_{\mu_1....\mu_r} \,=\, i m \zeta_{\mu_1....\mu_r} \,\, .
\ee

~

~



\section{The Yang-Mills-Chern-Simons theory and topological mass mechanism}

A remarkable no-go theorem on the topological mass mechanism in $d=4$ dimensions
has been presented in ref. \cite{h}. The argument is based on the non-existence of power-counting renormalizable gauge theories, constructed by consistent \emph{deformations} from the Cremmer-Scherk-Kalb-Ramond Abelian theory into a non-Abelian one. However, it
is interesting to analyze this theory in view of the gauge group
structure clarified here. Since an important ingredient of this negative result is
the impossibility of closing the algebra, one can hope that the present well defined
 group structure might solve these obstacles.

As shown above, the Cremmer-Scherk-Kalb-Ramond model may be
rigorously defined in DF (and generalized to non-Abelian gauge symmetry) as a Yang-Mills-Chern-Simons
theory (Ecs. (\ref{CS}),(\ref{YM})) :
\be\label{YMCS}
S({\cal A}) := m {\cal S}_{CS} +
{\cal S}_{YM}\, ,
\ee
where $m$ is a relative constant which has mass unit.
The structure of this theory is indeed completely similar to YMCS in $2+1$ dimensions, which are known to be finite
\cite{helpi}.
In this model the $\b$-symmetry is explicitly broken by the last term, however it remains totally invariant under the ordinary gauge symmetry $g =\exp \, i\, \a^a \tau^a$. In this sense this could be thought as a partially topologically massive model.

Because of that, we propose the action (\ref{YMCS}) in $d=4$ as a candidate to a well defined model with topological mass mechanism. This action results a non-Abelian BF one, which reads as
\be\label{YMCS-model}
{\cal S}(A , B) \equiv   m
\;\int\; Tr \; \left( [ A \wedge d {
B} + { B} \wedge d  A ] -  i 2 [A \wedge A\wedge
{ B}] \right)\,+ \,\,\int\,Tr\left(\,F\wedge{}^{*}F \,- \,{}^{*}H \wedge H\right)\;,
\ee
  where ${\cal A}=(A_\mu , B_{\mu \nu })$. Notice that a pure Chern-Simons boundary term could be added in a gauge invariant way, according to expression (\ref{CS4}), but we are not going to consider it here.

The kinetic $H^2$-term that provides the propagator for field $B$ is not $\beta$-invariant; but there is no other one, quadratic in the fields $A,B$ or their derivatives, being invariant under the most general combined $\alpha \oplus \beta$-tranformations (\ref{tr-gauge}).


\subsection{Propagators}

To compute the propagators for the
gauge-field excitations, we need to concentrate on the bosonic Lagrangian in terms
of the physical fields  $A_\mu $ and $B_{\mu \nu }$.
For the sake of reading off these propagators, we refer to the
bilinear sector of the Lagrangian, whose kinetic piece can be cast like below:
\begin{equation}
\begin{array}{ll}
{\cal L}_0 = \; Tr \;\left(\frac{1}{4} F_{\mu \nu} F^{\mu \nu} -
 \tilde G_{\mu} \tilde G^{\mu} + 2m A_{\mu}\tilde G^{\mu} \right), \label{kin}
\end{array}
\end{equation}

In order to invert the wave operator, we have to fix the gauge so as to make the matrix
 non-singular. This is accomplished by adding the gauge-fixing terms,

\begin{eqnarray}
{\cal L}_{A_\mu } &=&\frac 1{2\alpha }Tr(\partial _\mu A^\mu )^2; \\[0.3cm]
{\cal L}_{B_{\mu \nu }} &=&\frac 1{2\beta }Tr(\partial _\mu B^{\mu \nu
})^2.
\end{eqnarray}
where $\a , \b$ are constants. Although there is no gauge invariance under the sector $\beta$-gauge transformations, the last term is necessary to solve the equation of motion and find the propagators.

To read off the $A_\mu \ B_{\mu \nu}$-fields propagators, we shall use an extension of the spin-projection operator formalism
presented in \cite{Rivers,Nitsch}. The propagators we are looking for, were already computed in Ref. \cite{CSKR-prop} which may be written
 explicitly in terms of matrix elements,
\begin{equation}
<AA> = \frac{2i}{(4m^2+\Box)} \Theta_{\mu \nu} +
\frac{2 i \,\a}{\Box}
\Omega_{\mu \nu}\,\,\left(\sim k^{-2}\right) \,\,,  \label{propag1}
\end{equation}

\begin{equation}
<AB> =   -\frac{2m}{\Box} \frac{2i}{(4m^2+\Box)}
\epsilon^{\lambda \mu \nu }_{\alpha} \partial_{\lambda} \Theta^{\alpha k}\,\,\left(\sim k^{-3}\right) \,\,,
\end{equation}

\begin{equation}
<BB> = \frac{2i}{(4m^2+\Box)} (P^1_b)_{\alpha \beta \gamma k } +
\frac{i}{(16 m^{-1}- (8\b)^{-1}\Box )}(P_e^1)_{\alpha \beta \gamma k}\,\,\left(\sim k^{-2}\right) \,\,.
\label{propag2}
\end{equation}
which were expressed in terms of the algebra of projectors:

\begin{eqnarray}
(P^1_b)_{\mu \nu, \rho \sigma} &=& \frac{1}{2} (\Theta_{\mu \rho}\Theta_{\nu
\sigma} - \Theta_{\mu \sigma}\Theta_{\nu \rho}), \\
[0.3cm] (P^1_e)_{\mu \nu, \rho \sigma} &=& \frac{1}{2} (\Theta_{\mu
\rho}\Omega_{\nu \sigma} - \Theta_{\mu \sigma}\Omega_{\nu \rho} -\Theta_{\nu
\rho}\Omega_{\mu \sigma} + \Theta_{\nu \sigma}\Omega_{\mu \rho}),
\end{eqnarray}

\noindent
where $\Theta_{\mu \nu}$ and $\Omega_{\mu \nu}$ are, respectively, the
transverse and longitudinal projection operators, given by:

\begin{equation}
\Theta _{\mu \nu }=\eta _{\mu \nu }-\Omega _{\mu \nu },
\end{equation}

\noindent
and
\begin{equation}
\Omega_{\mu \nu} = \frac{\partial_{\mu} \partial_{\nu}}{\Box }.
\end{equation}
In Eq. (\ref{propag1}), we explicitly see that there is a massive mode for the gauge field $A_\mu$.

\vspace{0.3cm}

\noindent
The other operator coming from the Kalb-Ramond sector, $S_{\mu \gamma k}$, is
defined in terms of Levi-Civita tensor as

\begin{equation}
S_{\mu \gamma k} = \epsilon_{\lambda \mu \gamma k} \partial^{\lambda}.
\end{equation}

Let us write down the interaction terms that appear in the CS lagrangian:
\be
- m
\;\int\; Tr \; \left( [ A \wedge d {
B} + { B} \wedge d  A ] -  i 2 [A \wedge A\wedge
{ B}] \right)\,\;.
\ee
The first term is nothing but the third one in Eq. (\ref{kin}), then it is encoded in the propagators. The second term is the single CS vertex.


We also have the vertex terms from the YM sector (\ref{YM}):
\be
\, A\wedge A\wedge{}^{*}(A\wedge A) \,; \,2\,  \,{}^{*}(dA)\wedge(A\wedge A)\,;
 \,2\,\,{}^{*}(dB) \wedge (A\wedge B + B\wedge A ) \,; \,4\,{}^{*} (A\wedge B) \wedge B\wedge A ) \; .
\ee

From now on, all these terms, which will provide the different vertex diagrams, will be unambiguously referred by the respective subscripts $AAB, AAAA\equiv4A, AAA\equiv3A, ABB, AABB$.


\subsection{Power-counting rule for the primitive divergences}

we have
\be
\delta \equiv 2I_{AA} + 2I_{BB} + I_{AB} - 4(V-1) + V_{3A} + V_{BBA}
\ee
where $V=V_{AAB} + V_{3A} + V_{4A}+ V_{BBA}+ V_{AABB}$.
\be
 2I_{AA} + I_{AB} + E_A = 2V_{AAB} + 3V_{3A} + 4V_{4A}+V_{BBA}+ 2V_{AABB}
\ee
\be
 2I_{BB} + I_{AB} + E_B = V_{AAB} + 2V_{BBA}+ 2V_{AABB}
\ee
then we have
\be
\delta = 4-(I_{AB} + E_A + E_B + V_{AAB} )
\ee
which shows renormalizability for very generic diagrams, except for very special ones given by:
\be
(I_{AB} + E_A + E_B + V_{AAB} ) < 4
\ee
Sufficient number of external lines and/or external vertex, already guarantee that this last inequality is violated, and the diagram is then renormalizable.

\subsection{Discussion.}

By adopting an extended approach, referred to as tensor gauge field doublet formalism, we have found a possibility to build up the non-Abelian generalization of the CSKR model which opens up a viable path for a relevant topological mass mechanism in four dimensions. This construction favors a result that contrasts with a previous No-Go theorem \cite{h}, where is crucially argued that there is not a consistent deformation of the CSKR model.
We think that the objection raised in \cite{h} on the inconsistency of these type of models might be by-passed in the formalism presented here, since one assumption for the no-go result is that the non-Abelian extension of the Kalb-Ramond field, $B$ is uncharged. In mathematical language, this may be expressed by means of the transformation law:
\be
\delta B = D\b\equiv d\beta + [A, \b].
\ee
In our approach, however, which is based upon the doublet structure of the algebra, this brings up a novel term related precisely to the \emph{charge} of $B$, namely
\be
\delta B =  d\beta + [A, \b] + [B, \a].
\ee
The $[B, \a]$-term is genuinely due to our doublet formalism, and it plays an important role for the consistency of the theory.

\section{Concluding Comments}

In this work, we have presented a sort of extended algebraic structure and discussed some of its main tools in details, in order to build up general quantum field theories with $p$-tensors/forms as genuine gauge fields of conventional gauging methods associated to matter lagrangians.
A new topological invariant, shown to be connected to the Chern-Pontryagin index has been defined, and consequently, a 4d-Chern-Simons theory, involving both $A$ and $B$-fields was rigorously constructed.

We also have attempted at the construction of a consistent non-Abelian topologically massive gauge model in four dimensions and we argued it is a consistent quantum field-theoretic model. This can be considered to provide partially topological mass generation mechanism since it is gauge invariant under any conventional Lie group, but manifestly violates its extension to the $\b$-sector. The $\beta$-violating term $H^2$ in the YM sector is what provides the propagators for the theory; and there is no other one quadratic (second order) in the fields $(A, B)$ being invariant under the most general combined $\alpha \oplus \beta$-tranformations (\ref{tr-gauge}). This constitutes a sort of ``no-go'' result for this type of term in the present formalism. Therefore, one of our main results in this sense is precisely that the Chern-Simons action defined in Sec. 3 is the most general theory with this most general invariance in four dimensions. At this point, we can claim that the such full invariance can be realized at least for a topological action.

As a next step, we have to include the matter sector with the aim of reproducing/fiting the main features of the Standard Model. Another issue whose result will be reported in a forthcoming work concerns the Goldston theorem, which, in the context of the DF, presents interesting aspects and involves a spin-1 Goldstone boson along with the conventional scalar one whenever there occurs  spontaneous symmetry breaking. The role of the would-be spin-1 Goldstone bosons of the DF may be another problem to be understood in connection with the gravitational Higgs mechanism.

\section{Acknowledgements}

M.B.C. is grateful to CBPF for the warm hospitality during the periods of visit for the accomplishment of this paper. Special thanks are due to A. Lahiri for useful comments, and pointing out references.
This work was partially supported by: CONICET PIP 2010-0396, ANPCyT PICT 2007-0849, PCI-MCTI Grants.


\begin{thebibliography}{99}






\bibitem{kr0} M.~Kalb and P.~Ramond,
Phys.\ Rev.\ D {\bf 9}, 2273 (1974).

\bibitem{kr}
D. Z. Freedman, CALT-68-624. K.~Seo, M. Okawa and A. Sugamoto,

Phys.\ Rev.\ D {\bf 19}, 3744 (1979).

D. Z. Freedman and P. K. Townsend,
Nucl.\ Phys.\ B {\bf 177}, 282 (1981).

\bibitem{it5} Y. Nambu Phys.\ Rept.\  {\bf C23} (1976) 251.


\bibitem{it6} P. K. Townsend and P. van Nieuwenhuizen, Nucl.\ Phys.\
{\bf D120} (1977) 301.

\bibitem{aplic}
A. S. Schwarz,
Lett.\ Math.\ Phys.\  {\bf 2}, 247 (1978).

A. S. Schwarz,
Commun.\ Math.\ Phys.\  {\bf 67}, 1 (1979).

E. Witten,
Commun.\ Math.\ Phys.\  {\bf 117}, 353 (1988).

D. Birmingham, M. Blau, M. Rakowski and G. Thompson,
Phys.\ Rept.\  {\bf 209}, 129 (1991).

\bibitem{tm0}E. Cremmer and J. Scherk,
Nucl.\ Phys.\ B {\bf 72}, 117 (1974).

\bibitem{tm}
A. Aurilia and Y. Takahashi,
Prog.\ Theor.\ Phys.\  {\bf 66}, 693 (1981).

T. R. Govindarajan,
J.\ Phys.\ G {\bf 8}, L17 (1982).

T. J. Allen, M. J. Bowick and A. Lahiri,
Phys.\ Lett.\ B {\bf 237}, 47 (1990).

T. J. Allen, M. J. Bowick and A. Lahiri,
Mod.\ Phys.\ Lett.\ A {\bf 6}, 559 (1991).

\bibitem{la} A. Lahiri, Mod. Phys. Lett. A17 (2002) 1643;
A. Lahiri, J.Phys.A35 (2002) 8779, and references therein.

\bibitem{otro} P. G. O. Freund and R. I. Nepomechie, Nucl. Phys. B199 (1982)
482.

\bibitem{ultimo} C. Hofman, {\it Nonabelian 2 Forms},
 e-Print Archive: hep-th/0207017.

 \bibitem{h} M. Henneaux, S. P. Sorella, O. S. Ventura, C. Sasaki,
V. Lemos and L. C. Q. Vilar, {\em Phys. Lett.}{\bf B 410}(1997)
195.



\bibitem{helpi}
O. M. Del Cima, D. H. T. Franco, J. A. Helayel-Neto and O. Piguet,
Lett. Math. Phys. 47 (1999) 265.



\bibitem{0it} E. Di Grezia and S. Esposito {\it Minimal Coupling of the Kalb-Ramond
field to a Scalar Field}, hep-th/0304058.

\bibitem{it11}
S. Kar, P. Majumdar, S. SenGupta and A. Sinha,
Eur.\ Phys.\ J.\ C {\bf 23}, 357 (2002) [arXiv:gr-qc/0006097].

\bibitem{it12}
E. Gorbatov, V. S. Kaplunovsky, J. Sonnenschein, S. Theisen and S.
Yankielowicz,
JHEP {\bf 0205}, 015 (2002) [arXiv:hep-th/0108135].

\bibitem{it13} S. SenGupta and A. Sinha,
Phys.\ Lett.\ B {\bf 514}, 109 (2001) [arXiv:hep-th/0102073].

\bibitem{nunes-3-tensor}
T. S. Koivisto, N. J. Nunes, Phys. Lett. B 685 (2010) 105-109; Phys. Rev. D 80 (2009) 103509.


\bibitem{dkr} M. Botta Cantcheff,
{\it The Kalb-Ramond field as a connection on a flat space time},
hep-th/0212180.

\bibitem{krvec} M. Botta Cantcheff, {\it
Two form gauge field theories and ``no go" for Yang Mills
relativistic actions}, hep-th/0311183.

\bibitem{gkr} M. Botta Cantcheff, {\it
On the Group Structure of the Kalb-Ramond Gauge Symmetry.},
hep-th/0211239.


\bibitem{teit}  C. Teitelboim,
Phys. Lett. {\bf B 167} (1986) 63; M. Henneaux, C. Teitelboim,
Found. Phys. {\bf 16} (1986) 593 ; M. Henneaux, B. Knaepen,
 Phys. Rev. {\bf D 56} (1997) 6076,
and references therein.

\bibitem{DF} M Botta Cantcheff, \emph{Doublet groups, extended Lie algebras, and well defined gauge theories for the two form field}; e-Print: hep-th/0310156; Int. J. Mod. Phys. A20 (2005) 2673.

 \bibitem{matsuo-ho} Pei-Ming Ho, Y. Matsuo, \emph{ Note on non-Abelian two-form gauge fields}, [arXiv:1206.5643 [hep-th]]

\bibitem{dob} M. Botta Cantcheff,
Phys. Lett. B 533 (2002) 126 and Eur. Phys. Jour. C6 vol.4
(2002)1-14 (e-Print Archive: hep-th/0107123) ;  M. Botta Cantcheff
and J. A. Helay\"el-Neto, Phys. Rev. D67 (2003) 025016.

\bibitem{Rivers} R.J. Rivers, Nuovo Cimento. 34, 387, (1964).


\bibitem{Nitsch} R. Kuhfuss and J. Nitsch, Gen. Rel. Grav. 18, 1207, (1986).

\bibitem{CSKR-prop} C. N. Ferreira, M. B. D. S. M. Porto and J. A. Helay\"{e}l-Neto ;
Nucl. Phys. B 620 (2002) 181;
 H. R. Christiansen, M. S. Cunha, J. A. Helay\"{e}l-Neto,
 L. R. U. Manssur and A. L. M. A. Nogueira; Int. J. Mod. Phys. A 14 (1999) 1721;
W. A. Moura-Melo, N. Panza and J. A.
Helay\"{e}l-Neto; Int. J. Mod. Phys. A 14 (1999) 3949.

\bibitem{next} M. Botta Cantcheff and J. A. Helay\"el-Neto, work in progress.

\bibitem{luk} I.
Bandos, J. Lukierski, D. P. Sorokin, Phys. Rev. D61 (2000) 045002
(e-Print Archive: hep-th/9904109).

\bibitem{robinson} Sparling G A J 1999 Gen. Rel. Grav. 31 (1999) 837; Nurowski and D. C. Robinson, Class. Quantum Grav. 18 (2001) L81; Nurowski and D. C. Robinson, Class. Quantum Grav. 19 (2002) 2425.


\bibitem{asoc} R. D. Schafer, {\it An Introduction to
Non-Associative Algebras}, Pure And Applied Mathematics 22,
Academic Press, New York (1966); H. Carrion {\it Octonions and its
Applications in Physics}, PhD. Thesis (CBPF, june 2003), and
references therein.

\bibitem{nogue} A.L.M.A. Nogueira, M. Botta Cantcheff,  \textit{Gauge
Supersymmetry in the Doublets Formalism}; XXVI Brasilian National Meeting on Particle and Fields (SBF, Brasil), Oct 2005.


\bibitem{sav} G. Savvidy,  Proc. Steklov Inst. Math.272 (2011) 201; Phys. Lett. B 694 (2010).



\end{thebibliography}
\end{document}